\documentclass[prl,twocolumn,floatfix,
amsmath,amssymb,aps,superscriptaddress]{revtex4-1}
\usepackage{dcolumn,amsmath}
\usepackage[T2A]{fontenc}
\usepackage[utf8]{inputenc}
\usepackage{graphicx}
\graphicspath{ {./Images/} }
\usepackage{bm}
\usepackage{indentfirst} 
\usepackage{physics}

\begin{document}

\title{Effect of nuclear magnetization distribution within the Woods-Saxon model: Hyperfine splitting in neutral Tl}

\author{S. D. Prosnyak} \email{prosnyak\_sd@pnpi.nrcki.ru, prosnyak.sergey@yandex.ru}
\affiliation{Petersburg Nuclear Physics Institute named by B.P. Konstantinov of National Research Centre
``Kurchatov Institute'', Gatchina, Leningrad District 188300, Russia}
\affiliation{Saint Petersburg State University, 7/9
Universitetskaya nab., St. Petersburg, 199034 Russia}
\author{L. V. Skripnikov}\email{skripnikov\_lv@pnpi.nrcki.ru, leonidos239@gmail.com}
\affiliation{Petersburg Nuclear Physics Institute named by B.P. Konstantinov of National Research Centre
``Kurchatov Institute'', Gatchina, Leningrad District 188300, Russia}
\affiliation{Saint Petersburg State University, 7/9
Universitetskaya Naberezhnaya, St. Petersburg, 199034 Russia}

\homepage{http://www.qchem.pnpi.spb.ru}

\date{23.03.2021}

\begin{abstract}
Three models of the nuclear magnetization distribution are applied to predict the hyperfine structure of the hydrogen-like heavy ions and neutral thallium atoms: the uniformly magnetized ball model and single-particle models for the valence nucleon with the uniform distribution and distribution determined by the Woods-Saxon potential. 
Results for the hydrogen-like ions are in excellent agreement with  previous studies. 
The application of the Woods-Saxon model is now extended to the neutral systems with the explicit treatment of the electron correlation effects within the relativistic coupled cluster theory using the Dirac-Coulomb Hamiltonian. 
We estimate the uncertainty for the ratio of magnetic anomalies and numerically confirm its near nuclear-model independence. 
The ratio is used as a theoretical input to predict the nuclear magnetic moments of short-lived thallium isotopes. We also show that the differential magnetic anomalies are strongly model-dependent. 
The accuracy of the single-particle models significantly surpasses the accuracy of the simplest uniformly magnetized ball model for the prediction of this quantity. 
Skripnikov [L.V. Skripnikov, J. Chem. Phys. \textbf{153}, 114114 (2020)] has shown that the Bohr-Weisskopf contribution to the magnetic dipole hyperfine structure constant for an atom or a molecule induced by a heavy nucleus can be factorized into the electronic part and the universal nuclear magnetization dependent part. 
We numerically confirm this factorization for the Woods-Saxon single-particle model with an uncertainty less than 1\%.
\end{abstract}

\maketitle

\section{Introduction}
The hyperfine splitting in atomic spectra is of great interest for many physical applications. 
From a comparison of the theoretical and experimental values of the hyperfine structure (HFS) constants, one can test the accuracy of the electronic structure methods for atoms~\cite{Safronova:18,Porsev:2009,ginges2017ground,Skripnikov:2020b,GFreview} and molecules~\cite{KL95,Quiney:98,Titov:06amin,Skripnikov:15b,Skripnikov:15a,Sunaga:16,Fleig:17,Borschevsky:2020,Skripnikov:2020e}. 
Such electronic calculations are necessary to extract the value of the electric dipole moment of the electron and other fundamental constants and properties from the experimental data ~\cite{GFreview,Safronova:18,Skripnikov:14c,Skripnikov:17c,Skripnikov:15c,Skripnikov:17a}. 
Using the results of calculations and experimental data, it is possible to obtain the magnetic moments of short-lived nuclei ~\cite{Persson1998,0954-3899-37-11-113101,Schmidt:2018,barzakh2012hyperfine,Prosnyak:2020, Ginges:2020}. 
The obtained values can be used for the development of the nuclear structure theory.

In order to reproduce the experimental results for hyperfine splitting with an uncertainty of an order of 1\%, it is necessary to take into account both the finite charge distribution over the nucleus, the Breit-Rosenthal (BR) effect ~\cite{rosenthal1932isotope, crawford1949mf}, and the finite nuclear magnetization distribution, the Bohr-Weisskopf (BW) effect~\cite{bohr1950influence, bohr1951bohr, sliv1951uchet}. 
In studies of neutral atoms, the uniformly magnetized ball model is widely used to calculate the BW correction~\cite{Sapirstein:2003,konovalova2017calculation,Prosnyak:2020, Ginges:2018,kozlov2001parity}. 
The only parameter of this model is the radius of the ball, $R_M$.
Therefore, it can be not equal to the charge radius to reproduce the experimental value of the hyperfine splitting~\cite{elizarov2005hyperfine, Prosnyak:2020}, which raises questions about the physical meaning of such a model.

In this paper we study a more accurate and more physical single-particle (SP) nuclear magnetization distribution model. 
In this model it is assumed that the nuclear magnetic moment is induced by one unpaired nucleon, which has both the orbital motion and the spin.
We consider two approximations for the density of the unpaired nucleon. 
In the Woods-Saxon (WS) single-particle model, the wavefunction of this nucleon is obtained as a solution of the Schr{\"o}dinger equation with the WS potential~\cite{woods1954diffuse}. 
In the second single-particle model the uniform distribution (UD) of the valence nucleon is assumed. 
In the case of zero orbital momentum of the valence nucleon, this model is equivalent to the model of the uniformly magnetized ball.

For a point nuclear model, the ratio of the hyperfine splittings of two different isotopes 1 and 2 is proportional to the ratio of the nuclear \emph{g}-factors of the isotopes. 
However, this is not the case for the finite-size nucleus model due to the BR and BW effects. 
The corresponding correction, ${}^{1}\Delta^{2}$, is called the nuclear magnetic hyperfine anomaly:
\begin{equation}
\label{anomalyEq}
{}^{1}\Delta^{2}=\frac{A_1 g_2}{A_2 g_1}-1,
\end{equation}
where $A_1$ and $A_2$ are HFS constants [see Eq.~(\ref{hfs_el}) below] for a given electronic state, $g_1$ and $g_2$ are the nuclear \emph{g}-factors of the considered isotopes $1$ and $2$. 
The ratio of magnetic anomalies is a key theoretical input to obtain the magnetic moments of short-lived isotopes~\cite{Persson1998,0954-3899-37-11-113101,Schmidt:2018,barzakh2012hyperfine}.

In the present paper we apply the WS model to predict the contribution of the BW effect to the hyperfine structure of the neutral Tl atom in the ground and the first excited electronic states. 
As far as we know, this model has not been previously used to calculate the hyperfine structure of the neutral thallium atom with the explicit and direct treatment of the electron correlation effects. 
Results are compared with the values obtained within the uniformly magnetized ball model~\cite{Prosnyak:2020}. 
Next we compare predictions for the ratio of hyperfine magnetic anomalies and the differential hyperfine anomaly within different models. 
For the former we verify its near model independence and use it to deduce the magnetic moment values of the short-lived isotopes of thallium. 
For the latter we show that the SP models give results far better than those of the simple uniformly magnetized ball model. 
Finally, we numerically check the factorization of the BW contribution into the electronic and nuclear magnetization distribution dependent parts, introduced in Ref.~\cite{Skripnikov:2020e} for the WS model.

\section{Theory}
In the point magnetic dipole approximation the HFS constant $A^{(0)}$ for the atomic electronic state $\Psi_{JM_J}$ with the total electronic momentum $J$ and its projection $M_J$ on the axis $z$ can be calculated using the following expression:
\begin{equation}
    \label{hfs_el}
    A^{(0)} = \frac{\mu}{I M_J}\langle \Psi_{JM_J} |\frac{[\mathbf{r}_{el}\times \bm{\alpha}]_z}{r_{el}^3} | \Psi_{JM_J} \rangle,
\end{equation}
where $\mu$ is the value of the nuclear magnetic dipole moment, $I$ is the nuclear spin, $\bm{\alpha}$ are Dirac matrices, and $\mathbf{r}_{el}$ is the electron radius vector. 
The electronic wavefunction $\Psi_{JM_J}$ is calculated assuming the finite nuclear charge distribution. 
This means that the Breit-Rosenthal effect is considered nonperturbatively and is included in $A^{(0)}$.
In this case, the expression for the hyperfine splitting constant has the following form:
\begin{equation}
    A = A^{(0)} - A^{\rm BW}+A^{\rm QED} = A^{(0)}(1-\varepsilon)+A^{\rm QED},
\label{AA0}    
\end{equation}
where $A^{\rm BW}$ is the Bohr-Weiskopf contribution, $\varepsilon$ is the relative Bohr-Weiskopf correction and $A^{\rm QED}$ is the QED contribution. 
Below we do not consider the $A^{\rm QED}$ term in calculations of neutral systems.

In this paper we consider the SP nuclear magnetization distribution models in which the nuclear magnetization is generated by a single valence nucleon. 
In the WS model of the nucleus, the wavefunction of the valence nucleon is determined as a solution of the Schr{\"o}dinger equation with the Woods-Saxon potential
\cite{woods1954diffuse, rost1968proton}:
\begin{equation}
    U(r) = V(r) + V_C(r) + V_{SO}(r),
\end{equation}
where
\begin{equation}
    V(r) = -\frac{V_0}{1+e^{(r-R_0)/a}},
\end{equation}
\begin{equation}
    \begin{matrix}
    V_{C}(r)& =
    & \left\{
    \begin{matrix}
    (Z-1)/r & r \geqslant  R_C \\
    (Z-1)  (3-r^{2}/R^{2}_C) /2R_C & r \leqslant  R_C
    \end{matrix} \right.
    \end{matrix},
\end{equation}
\begin{equation}
    V_{SO}(r) 
    = \lambda (\frac{\hslash}{2 m_p c})^2 \frac{1}{r} \frac{d}{dr}\frac{V_0}{1+e^{(r-R_{SO})/a}} \: \pmb{ \sigma }\cdot \pmb{l}.
\end{equation}
Here $R_C=\sqrt{5/3}\langle r_c^2\rangle^{1/2}$ is the nuclear charge radius and $\langle r_c^2\rangle^{1/2}$ is the rms charge radius.
Parameters of the WS potential $R_0$, $R_{SO}$, $a$, $V_0$, and $\lambda$, are listed in Table~\ref{table:1}.
If the valence nucleon is the neutron then the Coulomb term $V_C$ should be omitted. 
\begin{table}[ht]
\caption{Parameters of the WS potential from Ref.~\cite{rost1968proton}. 
The radii are calculated as $R_0 = r_0 A^{1/3}$ and $R_{SO} = r_{SO} A^{1/3}$, where $A$ is the mass number.}
\centering
\begin{tabular}{lccccr}
\hline
\hline
 & $r_0$ (fm) & $r_{SO}$ (fm) & $a$ (fm)& $V_0$ (MeV)& $\lambda$\\
\hline
Proton & 1.275 & 0.932 & 0.70 & 58.7 & 17.8\\
Neutron & 1.347 & 1.280 & 0.70 & 40.6 & 31.5\\
\hline 
\hline
\end{tabular}
\label{table:1}
\end{table}

The BW correction $\varepsilon$ can be written as follows
\cite{bohr1950influence, bohr1951bohr, le1963hyperfine}:
\begin{eqnarray}
    \varepsilon = \frac{g_S}{g_I}\bigg[\frac{1}{2I}\langle K_S \rangle+\frac{(2I-1)}{8I(I+1)}\langle K_S - K_L\rangle\bigg]\nonumber\\ + \frac{g_L}{g_I}\bigg[\frac{(2I-1)}{2I}\langle K_L \rangle+\frac{(2I+1)}{4I(I+1)} \frac{m_p}{\hbar^2} \langle \phi_{SO}r^2 K_L\rangle\bigg]
\label{eps_1}
\end{eqnarray}
for $I=L+1/2$, and
\begin{eqnarray}
    \varepsilon = \frac{g_S}{g_I}\bigg[-\frac{1}{2(I+1)}\langle K_S \rangle-\frac{(2I+3)}{8I(I+1)}\langle K_S - K_L\rangle\bigg]\nonumber\\ + \frac{g_L}{g_I}\bigg[\frac{(2I+3)}{2(I+1)}\langle K_L \rangle-\frac{(2I+1)}{4I(I+1)} \frac{m_p}{\hbar^2} \langle \phi_{SO}r^2 K_L\rangle\bigg]
\label{eps_2}
\end{eqnarray}
for $I=L-1/2$.
Here $\phi_{SO}$ is the radial part of the spin–orbit interaction $V_{SO} = \phi_{SO} \: \pmb{ \sigma }\cdot \pmb{l}$, and $g_I$ is the $g$ factor of the considered nucleus. 
For the valence proton we set $g_L=1$, for the valence neutron we set $g_L=0$. 
$g_S$ is obtained from the following equations:
\begin{equation}
    \frac{\mu}{\mu_N} = \frac{1}{2}g_S + \bigg[I-\frac{1}{2}+\frac{2I+1}{4(I+1)}\frac{m_p}{\hbar^2}\langle \phi_{SO}r^2\rangle\bigg]g_L
\label{gsLm}         
\end{equation}
for $I=L+1/2$, and
\begin{equation}
    \frac{\mu}{\mu_N} = -\frac{I}{2(I+1)}g_S + \bigg[\frac{I(2I+3)}{2(I+1)}-\frac{2I+1}{4(I+1)}\frac{m_p}{\hbar^2}\langle \phi_{SO}r^2\rangle\bigg]g_L
\label{gsLp}      
\end{equation}
for $I=L-1/2$.
$\langle K_S \rangle$ and $\langle K_L \rangle$ are obtained by averaging functions $K_S(r)$ and $K_L(r)$ over the density of the valence nucleon $\left|u(r)\right|^{2}$:
\begin{equation}
    \langle K_{S,L} \rangle = \int_{0}^{\infty} K_{S,L}(r)  \left | u(r) \right |^{2}  r^2 \, dr.  
\end{equation}

Functions $K_S(r)$ and $K_L(r)$ in the case of a hydrogenlike ion have the following form:
\begin{equation}
K_S(r) = \dfrac{\displaystyle\int_{0}^{r}{f g \, dr_{el}} }{\displaystyle\int_{0}^{\infty}{f g \, dr_{el}}},
\label{K_S}
\end{equation}
\begin{equation}
K_L(r) = \dfrac{\displaystyle\int_{0}^{r}{(1-r_{el}^3/r^3) f g \, dr_{el}} }{\displaystyle\int_{0}^{\infty}{f g \, dr_{el}}},
\label{K_L}
\end{equation}
where $g$ and $f$ are the radial parts of the Dirac wavefunction of the electron. 
For the $1s$ ground state of the hydrogen-like ion, the following approximate expressions can be used
~\cite{bohr1950influence, shabaev1994hyperfine}
\begin{equation}
K_S(r) = b \bigg[\frac{a_1}{2}\Big(\frac{r}{R_C}\Big)^2+
\frac{a_2}{4}\Big(\frac{r}{R_C}\Big)^4+
\frac{a_3}{6}\Big(\frac{r}{R_C}\Big)^6\bigg],
\label{K_S_app}
\end{equation}
\begin{equation}
K_L(r) = 3b \bigg[\frac{a_1}{10}\Big(\frac{r}{R_C}\Big)^2+
\frac{a_2}{28}\Big(\frac{r}{R_C}\Big)^4+
\frac{a_3}{54}\Big(\frac{r}{R_C}\Big)^6\bigg].
\label{K_L_app}
\end{equation}
The expansion coefficients $b$ and $a_i$ can be found in Ref.~\cite{shabaev1994hyperfine}.

In the approximation of a uniformly distributed valence nucleon, the density of the valence nucleon has the following form:
\begin{equation}
    \left | u(r) \right |^{2} = \frac{3}{R_C^3}\theta(R_C - r),
\end{equation}
where $\theta(R_C - r)$ is the Heaviside step function:
\begin{equation}
    \theta(R_C - r) = \begin{cases} 1, & \mbox{if } r< R_C; \\ 0, & \mbox{if } r > R_C . \end{cases} 
\end{equation}
Note that for this model the terms with the spin-orbit interaction in Eqs.~(\ref{eps_1}) and (\ref{eps_2}) should be omitted.

Hyperfine magnetic anomalies (\ref{anomalyEq}) can be used to determine the magnetic moments of short-lived isotopes \cite{0954-3899-37-11-113101,Schmidt:2018,barzakh2012hyperfine,Ginges:2020}. 
We denote stable and short-lived isotopes by 1 and 2, respectively. 
Using the experimentally obtained HFS constants $A_1$ and $A_2$ for a given electronic state $b$, the magnetic moment of the stable isotope, $\mu_1$, and the hyperfine magnetic anomaly, one can determine the magnetic moment $\mu_2$ of the short-lived isotope:
\begin{equation}
\label{mu2}
\mu_2=\mu_1\cdot\frac{A_2[b]}{A_1[b]}\cdot\frac{I_2}{I_1}
\cdot(1+{}^{1}\Delta^{2}[b]).
\end{equation}
A direct calculation of the anomaly ${}^{1}\Delta^{2}[b]$ is quite difficult due to a strong dependence of the result on the choice of the nuclear model. 
However, the ratio of the anomalies
\begin{equation}
\label{rohfa}
{}^1k^2[a,b]={}^{1}\Delta^{2}[a]/{}^{1}\Delta^{2}[b] 
\end{equation}
for two electronic states $a$ and $b$ turns out to be fairly stable, which we verify below. 
Using this fact, it is possible to extract the desired nuclear magnetic moment of a short-lived isotope. 
For this, it is necessary to know the magnetic moment of a stable isotope, as well as the hyperfine constants $A_{1,2}[a]$ and $A_{1,2}[b]$ for the electronic states $a$ and $b$ of the nuclei under consideration. 
For convenience, we introduce the so-called differential hyperfine magnetic anomaly $^1\theta^2[a,b]$~\cite{Persson1998,barzakh2012hyperfine}:
\begin{equation}
\label{theta}
    {}^1\theta^2[a,b]=
    \frac{A_1[a]}{A_2[b]}\frac{A_2[a]}{A_1[b]} - 1=
    \frac{1+{}^1\Delta^2[a]}{1+{}^1\Delta^2[b]} - 1.
\end{equation}
The important feature of $^1\theta^2[a,b]$ is that it is independent of the magnetic moments and spins of the nuclei under consideration.
As it can be seen from Eq.~(\ref{theta}), $^1\theta^2[a,b]$ can be determined using only the experimental values of the hyperfine constants. 
Substituting the ratio of hyperfine magnetic anomalies into Eq.~(\ref{theta}), we find~\cite{0954-3899-37-11-113101,Schmidt:2018,barzakh2012hyperfine}:
\begin{equation}
{}^{1}\Delta^{2}[b]=\frac{{}^1\theta^2[a,b]}
   {^1k^2[a,b]-{}^1\theta^2[a,b]-1}.
\label{DeltaB}   
\end{equation}
One can put ${}^{1}\Delta^{2}[b]$ into Eq.~(\ref{mu2}) to finally obtain the desired nuclear magnetic moment. 
Below we explore the model dependence of both the ratio of hyperfine magnetic anomalies $^1k^2[a,b]$ and the differential magnetic anomaly ${}^1\theta^2[a,b]$.

\section{Calculation details}
The values of the charge radii of the stable nuclei were taken from Ref.~\cite{ANGELI201369}.
The charge radii of the short-lived thallium isotopes were taken from Ref.~\cite{barzakh2013changes}. 
Nuclear magnetic moments listed in Table \ref{table:2} were taken from Ref.~\cite{stone2014table} for stable nuclei and Ref.~\cite{Prosnyak:2020} for short-lived thallium isotopes.
\begin{table}[ht]
\caption{Employed parameters of the nuclei: valence nucleon state, nuclear magnetic dipole moments~\cite{stone2014table,Prosnyak:2020,Skripnikov:18a,Skripnikov:2020a,antuvsek2020nmr} and charge radii~\cite{ANGELI201369, barzakh2013changes}. 
In the square brackets the values of the magnetic moments used in previous papers are given; they have been revisited in recent papers~\cite{Skripnikov:18a,Skripnikov:2020a,antuvsek2020nmr}.}
\centering
\begin{tabular}{lccc}
\hline
\hline
Nucleus & State & $\mu_I/\mu_N$ & $\langle r_c^2\rangle^{1/2}$ (fm) \\
\hline
${}^{185}{\rm Re}$      & $2d_{5/2}$ & 3.1570(29) [+3.1871(3)] & 5.3596 \\
${}^{191}$Tl$^m$        & $1h_{9/2}$ & 3.79(2) & 5.4310 \\
${}^{193}$Tl$^m$        & $1h_{9/2}$ & 3.84(3) & 5.4382 \\
${}^{203}{\rm Tl}$      & $3s_{1/2}$ & 1.62225787(12) & 5.4666 \\
${}^{205}{\rm Tl}$      & $3s_{1/2}$ & 1.63821461(12) & 5.4759 \\
${}^{207}{\rm Pb}$      & $3p_{1/2}$ & 0.59102(18) [0.592583(9)] & 5.4943 \\
${}^{209}{\rm Bi}$      & $1h_{9/2}$ & 4.092(2) [4.1106(2)] & 5.5211 \\
\hline 
\hline
\end{tabular}
\label{table:2}
\end{table}
WS potential parameters were taken from Ref.~\cite{rost1968proton} and are listed in Table~\ref{table:1}.

\begin{figure}[ht]
\begin{center}
\includegraphics[width=1.0\linewidth]{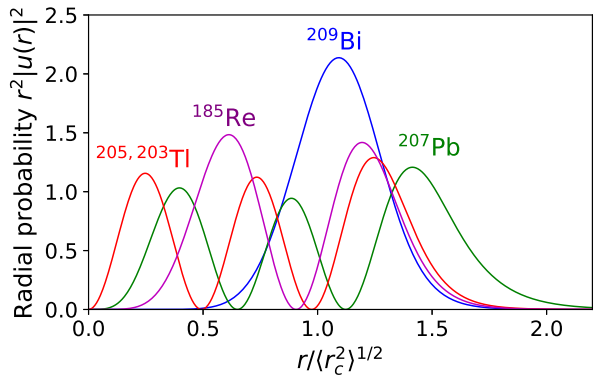}
\end{center}
\caption{Calculated radial probability densities of the valence nucleon for various nuclei. 
The densities for ${}^{203}$Tl and ${}^{205}$Tl isotopes coincide with rather high accuracy and are indicated by a single line.}
\label{figure:1}
\end{figure}

To obtain the nucleon wave function in the WS model the radial Schr{\"o}dinger equation has been solved on the grid using the code developed in the present paper. 
Calculated radial probability densities of a valence nucleon for different isotopes are shown in Fig.~\ref{figure:1}.
The electronic wavefunction for the hydrogen-like ions have been obtained by the numerical solution of the Dirac equation using the Gaussian-type basis set. 
This basis set includes $50$ $s-$type functions, with exponential parameters forming a geometric progression. 
The common ratio of this progression is 1.8, and the largest element is $5\cdot 10^8$.

In HFS calculations of the neutral thallium atom, the QED effects were not taken into account. 
Atomic orbitals for subsequent correlation calculations were obtained using the Dirac-Hartree-Fock (DHF) method, where the Fock operator is determined by averaging electronic shell configurations over $6p_{j=1/2}^1$ and $6p_{j=3/2}^1$ for $6P_{1/2}$ and $6P_{3/2}$ electronic states. 
For the $7S_{1/2}$ electronic state the averaging has been performed over the $7s_{j=1/2}^1$ configuration. 
The main correlation calculations that include all 81 electrons have been performed using the coupled cluster method with single, double, and perturbative triple amplitudes, CCSD(T)~\cite{Visscher:1996,bartlett2007coupled} within the Dirac-Coulomb Hamiltonian. 
In these calculations the uncontracted Dyall's AAE4Z basis set~\cite{Dyall:12} augmented with one $h-$ and one $i-$ type functions was used. 
It includes$35s-$, $32p-$, $22d-$, $16f-$, $10g-$, $5h-$ and $2i-$ type functions. 
For the calculation virtual orbitals were truncated at the energy of 10000 hartree. 
The importance of the high energy cutoff for properties dependent on the behavior of the wavefunction close to the heavy-atom nucleus has been demonstrated in Refs.~\cite{Skripnikov:17a,Skripnikov:15a}. 
In the tables below we also include corrections on the basis set size extension, high-order correlation effects beyond the CCSD(T) level and the Gaunt interaction contribution from Ref.~\cite{Prosnyak:2020}. 
The basis set correction has been calculated within the CCSD(T) method using the extended basis set that includes $44s$, $40p$, $31d$, $24f$, $15g$, $9h$ and $8i$ basis functions. 
$1s-3d$ electrons were excluded from the correlation treatment and the virtual orbitals were truncated at the energy of 150 Hartree in these calculations. 
Calculations of the contributions of correlation effects beyond the CCSD(T) model have been performed within the coupled cluster with single, double, triple, and perturbative quadruple amplitudes [CCSDT(Q)] method~\cite{Kallay:6,Kallay:1,Kallay:2}. 
In these calculations we have used the SBas basis set that consists of $30s-$, $26p-$, $15d-$, $9f-$ type functions and corresponds to the Dyall's CVDZ~\cite{Dyall:06,Dyall:98} basis set augmented by diffuse functions. 
As in the case of the basis set correction calculation, $1s-3d$ electrons were excluded from the correlation treatments. 
The contribution of the Gaunt interaction has been calculated within the SBas basis set using the CCSD(T) method. 
In this calculation, all electrons were correlated and all virtual orbitals within a given basis set were considered. 
Correlation calculations have been performed using the finite-field technique. For relativistic coupled cluster calculations the {\sc dirac15}~\cite{DIRAC15} and {\sc mrcc} codes~\cite{MRCC2020,Kallay:1,Kallay:2} were used. 
The code developed in Ref.~\cite{Skripnikov:16b} was used to calculate the HFS integrals in the approximation of a point magnetic dipole. 
The code for calculating the BW matrix elements in the WS model has been developed in the present paper.

\begin{table*}
\caption{Calculated values of the BW correction $\varepsilon$ (in \%) using the WS model for various hydrogen-like ions in the ground electronic state $1s$.}
\centering
\begin{tabular*}{0.77\linewidth}{lccccc}
\hline
\hline
Author, reference & ${}^{185}{\rm Re}^{74+}$ & ${}^{203}{\rm Tl}^{80+}$ & ${}^{205}{\rm Tl}^{80+}$ & ${}^{207}{\rm Pb}^{81+}$ & ${}^{209}{\rm Bi}^{82+}$\\
\hline
Shabaev \textit{et al.} \cite{shabaev1997ground}, Eqs. (\ref{K_S_app}), (\ref{K_L_app}), without SO & 1.20 & 1.77 & 1.77 & 4.19 & 1.33\\
Shabaev \textit{et al.} \cite{shabaev1997ground}, Eqs. (\ref{K_S_app}), (\ref{K_L_app}), with SO & 1.22 & 1.79 & 1.79 & - & 1.18\\
Gustavsson \textit{et al.} \cite{gustavsson2000thallium} & 1.18 & 1.74 & 1.74 & 4.29 & 1.31\\
This work, Eqs. (\ref{K_S_app}), (\ref{K_L_app}), without SO & 1.20 & 1.78 & 1.78 & 4.44 & 1.29\\
This work, Eqs. (\ref{K_S_app}), (\ref{K_L_app}), with SO & 1.22 & 1.80 & 1.79 & 4.47 & 1.17\\
This work, Eqs. (\ref{K_S}), (\ref{K_L}), without SO & 1.30 & 1.87 & 1.87 & 4.43 & 1.43\\
This work, Eqs. (\ref{K_S}), (\ref{K_L}), with SO & 1.32 & 1.89 & 1.89 & 4.45 & 1.30\\
\hline
Experiment & 1.35 & 2.21 & 2.23 & 3.81 & 1.02\\
\hline 
\hline
\end{tabular*}
\label{table:3}
\end{table*}

\section{Results and discussion}
To test the developed approach, the HFS constants of hydrogen-like ions were calculated. 
The obtained values are given in Table~\ref{table:3} and compared with the previous studies~\cite{shabaev1997ground, gustavsson2000thallium}. 
A slight difference between the present and the previous results can be explained by a different nuclear charge model. 
In the present calculations the Gaussian charge distribution model~\cite{Visscher:1997} was used, while in the previous calculations the Fermi distribution was employed. 
The Gaussian charge distribution model is widely used in the molecular calculations of HFS. 

Table~\ref{table:3} contains also the BW correction extracted from the experimental values of the HFS constants $A^{\rm exp}$~\cite{beiersdorfer2001hyperfine, beiersdorfer2003hyperfine} using the following expression: 
\begin{equation}
    \varepsilon_{\rm exp} = 1-\frac{(A^{\rm exp}-A^{\rm QED})}{\frac{\mu}{I\cdot M_J}\langle \Psi_{JM_J} |\frac{[\mathbf{r}_{el}\times \bm{\alpha}]_z}{r_{el}^3} | \Psi_{JM_J} \rangle}.
\end{equation}
For calculation of the denominator, we used the data from Ref.~\cite{shabaev1997ground} and the latest values of the nuclear magnetic moments. QED contributions $A^{\rm QED}$ were taken from Refs.~\cite{shabaev1997ground,shabaevpriv,artemyev2001vacuum}. 
Note that there is a small dependence of the BW correction calculated in the SP models due to the dependence of the parameter $g_S$ on the magnetic moment value [see Eqs.~(\ref{eps_1})-(\ref{gsLp})].
Therefore, to be able to compare with previous calculations of the BW correction for H-like ions we used the same values of the magnetic moments that have been used in the previous papers.
However, to obtain the $\varepsilon_{\rm exp}$ values the revisited nuclear magnetic moment values~\cite{Skripnikov:18a,Skripnikov:2020a,antuvsek2020nmr} have been used (see Table~\ref{table:2}). 
One can see from Table~\ref{table:3} that the simplified Eqs. (\ref{K_S_app}) and (\ref{K_L_app}) give very good approximation to the more accurate Eqs.~(\ref{K_S}) and (\ref{K_L}). 

Tables \ref{table:4} and \ref{table:5} give the values of calculated HFS constants for the neutral ${}^{205}$Tl atom in the ground electronic state $6 P_{1/2}$ and the first excited state $6P_{3/2}$, respectively. 
In the last column, the values of HFS constants with BW contributions calculated within the WS model of the nuclear magnetization distribution are given. 
They were obtained using Eqs. (\ref{K_S}) and (\ref{K_L}) for one-electron matrix elements. 
For comparison, we also provide results obtained within the point magnetic dipole approximation (the second column) and the uniformly magnetized ball model from Ref.~\cite{Prosnyak:2020} (the third column). 
One can see from Tables~\ref{table:4} and \ref{table:5} a reasonable agreement between the HFS constants calculated in the ball model and in the WS model for $^{205}$Tl.
The theoretical uncertainty of the electronic structure calculation in Ref. \cite{Prosnyak:2020} was estimated as 1\% for $6 P_{1/2}$ and about 10\% for $6 P_{3/2}$. 
One can see very good agreement of the theoretical prediction of the HFS constant for the $6 P_{1/2}$ state with the experimental value, $21310.835(5)$~MHz. 
A reasonable agreement between the theoretical value of the HFS constant for the $6 P_{3/2}$ state and the experimental value, $265.0383(1)$~MHz, is obtained. 
It can be noted that the WS model also predicts large relative BW correction for this state (see a detailed discussion in Ref.~\cite{Prosnyak:2020}).

\begin{table}[ht]
\caption{Calculated values of the HFS constant of the $6P_{1/2}$ state of ${}^{205}$Tl (in MHz) using different levels of electronic theory and nuclear models. 
The numbers in the first line in columns 2 and 3 indicate the ratio of the model magnetic radius and the charge radius $R_M/R_C$. 
The values of the BW contributions, $-A^{\rm BW}$, are given in parentheses.}
\begin{tabular}{lrrr}
\hline
\hline
Method & 0~\cite{Prosnyak:2020}& 1.0~\cite{Prosnyak:2020} & WS \\
\hline
DHF &~ 18805 & ~~~18681 &~ 18696 \\
{} & {} & (-124) & (-109)\\
CCSD & 21965 & 21807 &~ 21826 \\
{} & {} & (-158) & (-139)\\
CCSD(T) & 21524 & 21372 &~ 21390 \\
{} & {} & (-152) & (-134)\\
\hline
~+Basis corr. & -21 & -- & -- \\
~+CCSDT-CCSD(T) & +73 & -- &~ -- \\
~+\footnotesize{CCSDT(Q)-CCSDT} & -5 & -- &~ -- \\
~+Gaunt & -83 & -- &~ --\\
\hline
Total$^a$ & 21488 & 21337 &~ 21354 \\
\hline
\hline
\end{tabular}
\label{table:4}
\\
$^a$ Instead of missing corrections, the contributions calculated for the point magnetic dipole moment model given in the first column were used.
\end{table}

\begin{table}[ht]
\caption{Calculated values of the HFS constant of the $6P_{3/2}$ state of ${}^{205}$Tl (in MHz) using different levels of electronic theory and nuclear models. 
The numbers in the first line in columns 2 and 3 indicate the ratio of the model magnetic radius and the charge radius $R_M/R_C$. 
The values of the BW contributions, $-A^{\rm BW}$, are given in parentheses.}
\begin{tabular}{lrrr}
\hline
\hline
Method & 0~\cite{Prosnyak:2020} & 1.0~\cite{Prosnyak:2020} & WS \\
\hline
DHF &~ 1415 & 1415 &~ 1415 \\
CCSD & 6 & 40 &~ 36 \\
{} & {} & (+34) & (+30)\\
CCSD(T) & 244 & 273 &~ 269 \\
{} & {} & (+29) & (+25)\\
\hline
~+Basis corr. & +4 & -- & -- \\
~+CCSDT-CCSD(T) & -49 & -- &~ -- \\
~+\footnotesize{CCSDT(Q)-CCSDT} & +14 & -- &~ -- \\
~+Gaunt & +1 & -- &~ --\\
\hline
Total$^a$ & 214 & 243 &~ 239 \\
\hline
\hline
\end{tabular}
\label{table:5}
\\
$^a$ Instead of missing corrections, the contributions calculated for the point magnetic dipole moment model given in the first column were used.
\end{table}

Table \ref{table:6} presents the values of calculated ratios of hyperfine magnetic anomalies ${}^{205}k^x[7S_{1/2},6P_{1/2}]$, where $x$ is ${}^{203}$Tl, ${}^{193}$Tl$^m$ or ${}^{191}$Tl$^m$. 
Results are given at different levels of the electronic structure theory for three models of the magnetization distribution: the uniformly magnetized ball model~\cite{Prosnyak:2020} and the UD and WS single-particle models. 
In the former model the ball radius is equal to the charge radius. The obtained values are in fairly good agreement. 
This numerically justifies the assumed near model independence of such a ratio. 
Thus, a theoretical calculation of the ratio of hyperfine magnetic anomalies for a pair of electronic states can be used to determine the magnetic moments of short-lived isotopes. 
It should be noted that, for stable isotopes, the charge and magnetization distribution effects give comparable contributions to the anomalies, and hence to their ratio. 
However, for the case of isotopes having different states of the valence nucleon, the main contribution to the anomaly comes from the BW effect.

\begin{table}[ht]
\caption{The ratio of magnetic hyperfine anomalies ${}^{205}k^x[7S_{1/2},6P_{1/2}]$, where $x$ is ${}^{203}$Tl, ${}^{193}$Tl$^m$ or ${}^{191}$Tl$^m$. For the Ball and UD models the magnetic rms radius was set to be equal the experimental rms charge radius.}
\centering
\begin{tabular}{lcccr}
\hline
\hline
 Nucleus & Method & Ball~\cite{Prosnyak:2020} & UD & ~~~WS \\
\hline
 {} & DHF & 3.77 & 3.77 & 3.85 \\
 ${}^{203}{\rm Tl}$ & CCSD & 3.38 & 3.38 & 3.44 \\
 {} & CCSD(T) & 3.47 & 3.47 & 3.54 \\
\hline 
 {} & DHF & 3.73 & 3.55 & 3.54 \\
 ${}^{193}$Tl$^m$ & CCSD & 3.36 & 3.23 & 3.22 \\
 {} & CCSD(T) & 3.45 & 3.32 & 3.31 \\
\hline
 {} & DHF & 3.74 & 3.55 & 3.54 \\
 ${}^{191}$Tl$^m$ & CCSD & 3.36 & 3.23 & 3.22 \\
 {} & CCSD(T) & 3.46 & 3.32 & 3.31 \\
\hline
\hline
\end{tabular}
\label{table:6}
\end{table}

Table \ref{table:7} gives the values of the differential hyperfine magnetic anomalies ${}^{205}\theta^{x}[7S_{1/2},6P_{1/2}]$ defined by Eq.~(\ref{theta}), where $x$ is ${}^{203}$Tl, ${}^{193}$Tl$^m$ or ${}^{191}$Tl$^m$. As in the previous case, three nuclear magnetization distribution models have been used: the simplest uniformly magnetized ball model and two single-particle models: UD and WS. The obtained values of the differential anomaly for ${}^{193}$Tl$^m$ and ${}^{191}$Tl$^m$ isotopes are slightly smaller than the estimate ${}^{205}\theta^{x(I=9/2)}[7S_{1/2},6P_{1/2}]=-1.2\cdot10^{-2}$ from Ref.~\cite{barzakh2012hyperfine}. This can be explained by the fact that in Ref.~\cite{barzakh2012hyperfine} the effective value of the orbital $g$ factor of the valence nucleon $g_L = 1.16$ from Ref.~\cite{Grossman:1999} has been used.
In our calculations, the value $g_L = 1.0$ has been used. 
For comparison, we have also performed calculations at the DHF level using the effective value of $g_L$ from paper~\cite{Grossman:1999} and the corresponding value $g_S$ derived using Eqs.~(\ref{gsLm}) and (\ref{gsLp}).
We estimate: ${}^{205}\theta^{x(I=9/2)}[7S_{1/2},6P_{1/2}]=-1.13\cdot10^{-2}$ and ${}^{205}\theta^{x(I=9/2)}[7S_{1/2},6P_{1/2}]=-0.95\cdot10^{-2}$ for the UD and WS models of magnetization distribution, respectively.

\begin{table}[ht]
\caption{The differential magnetic hyperfine anomalies ${}^{205}\theta^x[7S_{1/2},6P_{1/2}]$, where $x$ is ${}^{203}$Tl, ${}^{193}$Tl$^m$ or ${}^{191}$Tl$^m$, $10^{-4}$. 
For the ball and UD models the magnetic rms radius was set to be equal to the experimental rms charge radius.
The experimental values~\cite{barzakh2012hyperfine, chen2012absolute, lurio1956hfs} are given in the last column.}
\centering
\begin{tabular}{lccccr}
\hline
\hline
 Nucleus & Method & Ball~\cite{Prosnyak:2020} & UD & WS & Experiment\\
\hline
 {} & DHF & -1.09 & -1.09 & -0.86 & {}\\
 ${}^{203}{\rm Tl}$ & CCSD & -1.05 & -1.05 & -0.83 & -1.9(8)\\
 {} & CCSD(T) & -1.06 & -1.06 & -0.84 & {}\\
\hline 
 {} & DHF & -5.14 & -93 & -69 & {}\\
 ${}^{193}$Tl$^m$ & CCSD & -4.92 & -90 & -66 & -129(62)\\
 {} & CCSD(T) & -4.98 & -91 & -67 & {}\\
\hline
 {} & DHF & -6.06 & -96 & -72 & {}\\
 ${}^{191}$Tl$^m$ & CCSD & -5.80 & -92 & -69 & -154(60)\\
 {} & CCSD(T) & -5.87 & -93 & -70 & {}\\
\hline
\hline
\end{tabular}
\label{table:7}
\end{table}

As one can see from Table~\ref{table:7}, the dependence of the differential magnetic anomaly on the level of the included electronic correlation effects is slightly smaller than in the case of the ratio of the magnetic anomalies. 
In the case of the differential magnetic anomaly ${}^{205}\theta^{203}[7S_{1/2},6P_{1/2}]$, theoretical and experimental values are of the same order of magnitude. 
However, for the short-lived isotopes ${}^{193}$Tl$^m$ and ${}^{191}$Tl$^m$, SP models give much more accurate results than the model of a uniformly magnetized ball. 
This can be explained by the fact that the ${}^{205}$Tl and ${}^{203}$Tl thallium isotopes have the same valence nucleon state $s_{1/2}$ with zero orbital momentum (see Table~\ref{table:2}).
In this case the uniformly magnetized ball model reduces to the single-particle UD model. 
This is not the case for the short-lived isotopes ${}^{193}$Tl$^m$ and ${}^{191}$Tl$^m$ with the valence nucleon state having nonzero orbital momentum.
Thus, it follows from Table~\ref{table:7} that it is important to use nuclear magnetization distribution models that are more complex than the simplest uniformly magnetized ball model.

It has been shown in Ref.~\cite{Skripnikov:2020e} that the BW contribution to the hyperfine structure constant of an atom or a \textit{molecule} induced by a heavy nucleus can be factorized into the electronic part, $E$, and the universal nuclear magnetization distribution dependent part, $N$, with very high accuracy (see Eq. (29) in Ref.~\cite{Skripnikov:2020e}).
As it has been shown in Ref.~\cite{Skripnikov:2020e}, such factorization is valid for almost any electronic state and for calculations with the treatment of the electron correlation effects.
The electronic part depends only on the considered electronic state. The nuclear magnetization distribution dependent part does not depend on the actual electronic state. 
In Ref.~\cite{Skripnikov:2020e} the nuclear part corresponds to the matrix element of the BW correction operator over the $1s$ function of the corresponding hydrogenlike ion, $B_s$. 
In particular, it means that within a given level of the electronic structure theory the ratio of two BW corrections calculated using two different models of the nuclear magnetization distribution is equal to the ratio of the nuclear parts and should not be dependent on the level of the considered electronic structure theory.
Moreover, it should not be dependent on the actual electronic and charge state of the considered open-shell system (we do not consider here situations when the HFS constant is determined exclusively by an electron in the electronic state with $j\ge3/2$). 
Tables \ref{table:4} and \ref{table:5} give the BW contributions, $-A^{\rm BW}$, calculated within the uniformly magnetized ball model and the single-particle WS model for different levels of electronic structure theory (see the numbers in brackets). 
According to our findings, the ratio of these BW contributions is indeed practically (with the uncertainty less than 1\%) independent of the level of the electronic structure theory as well as of the considered electronic and charge state: $6P_{1/2}$ and $6P_{3/2}$ of the neutral Tl and $1S_{1/2}$ of the hydrogenlike Tl.

The theory formulated in Ref.~\cite{Skripnikov:2020e} can be also used to illustrate the dependence of the ratio of magnetic anomalies and the differential anomalies on the model of the nuclear magnetization distribution. 
For convenience of consideration, we rewrite Eq.~(\ref{AA0}) by separating further the Breit-Rosenthal correction $\delta$:
\begin{equation}
    A = A^{(0)}(1-\varepsilon) = A^{(p.n.)}(1-\delta)(1-\varepsilon),
\end{equation}
where $A^{(p.n.)}$ is the HFS constant corresponding to the point nucleus. 
In this case, in the leading order, the magnetic anomaly is determined by the magnetic and charge distribution contributions:
\begin{equation}
    {}^1\Delta^2 \approx {}^1\Delta^2_m + {}^1\Delta^2_c = \varepsilon_2 - \varepsilon_1 + \delta_2-\delta_1.
\end{equation}
For isotopes with different valence nucleon states the main contribution to the anomaly comes from the magnetic distribution term, while the charge distribution term, $(\delta_2-\delta_1)$, can be neglected for a qualitative treatment, i.e., ${}^1\Delta^2 \approx \varepsilon_2 - \varepsilon_1$. 
Using the factorization of the BW corrections~\cite{Skripnikov:2020e} we obtain the following expression:
\begin{equation}
    {}^1\Delta^2[a] \approx \varepsilon_2[a] - \varepsilon_1[a] = E[a](N_2 - N_1).
\end{equation}
As one can see, the ratio of anomalies for two electronic states depends on the ratio of electronic parts:
\begin{equation}
    {}^1 k ^2 [a, b] = \frac{{}^1\Delta^2[a]}{{}^1\Delta^2[b]} \approx \frac{E[a]}{E[b]}.
\end{equation}
A slight deviation from this equality can be due the neglected charge distribution contribution. 
Thus, for this case, it is reasonable to suggest that the uncertainty of the ratio of magnetic anomalies is mainly due to the uncertainty of the electronic structure calculation. 
For example, according to Table~\ref{table:6}, below we assume ${}^{205}k^{x(I=9/2)}[7S_{1/2},6P_{1/2}]=3.31(10)$. 
At the same time, a differential anomaly depends on both the electronic and nuclear parts:
\begin{equation}
    {}^1\theta^2 [a, b] \approx {}^1\Delta^2[a]-{}^1\Delta^2[b] = 
    (E[a] - E[b])(N_2-N_1).
\end{equation}

Table~\ref{table:8} gives the values of magnetic moments for short-lived thallium nuclei calculated according to Eqs.~(\ref{mu2})-(\ref{DeltaB}) using the calculated ratio of anomalies from Table~\ref{table:6} and the experimental values of HFS constants from Ref.~\cite{barzakh2012hyperfine}. 
For ${}^{193}\text{Tl}^m$ and ${}^{191}\text{Tl}^m$ isotopes, this ratio is the same within a given uncertainty. 
Therefore, the same value, 3.31(10), has been used for other isotopes in Table~\ref{table:8}, all of which also have one valence proton in the $1h_{9/2}$ state.
Following Ref.~\cite{barzakh2012hyperfine}, we used the mean weighted value of the experimental differential anomaly ${}^{205}\theta^{x(I=9/2)}[7S_{1/2},6P_{1/2}] = -1.53(37)\cdot10^{-2}$ for the isotopes under consideration. 
The magnetic moments obtained with this value are given in the third column of Table~\ref{table:8}. 
As one can see, the obtained values are in good agreement with the results of Ref.~\cite{barzakh2012hyperfine}. 
Their difference is mainly due to the different values of the ratio of the magnetic anomalies. 
In the present paper the WS model has been used while in Ref.~\cite{barzakh2012hyperfine} the single particle model with a uniform valence nucleon distribution model from Ref.~\cite{gustavsson2000thallium} has been used.
Alternatively, the differential anomaly can be determined for each isotope separately using Eq.~(\ref{theta}). 
For this, the experimental values of HFS constants $A_{205}[7S_{1/2}]=12296.1(7)$ from Ref.~\cite{chen2012absolute}, $A_{205}[6P_{1/2}]=21310.835(5)$ from Ref.~\cite{lurio1956hfs}, and the hyperfine constants for short-lived thallium isotopes from Ref.~\cite{barzakh2012hyperfine} were used. 
The obtained results are given in the last column of Table~\ref{table:8}. 
The determined magnetic moments are in good agreement with the values $\mu({}^{193}\text{Tl}^m) = 3.84(3)\mu_N$ and $\mu({}^{191}\text{Tl}^m) = 3.79(2)\mu_N$ from Ref.~\cite{Prosnyak:2020}, where the same approach was used.
The main source of the magnetic moments' uncertainty is the experimental uncertainty of the HFS constants of the short-lived isotopes. 
\begin{table}[ht]
\caption{Magnetic moments $\mu(\mu_N)$ for short-lived thallium isotopes with $I=9/2$.
The values in column 3 were obtained using the averaged value of the differential anomaly, while the values in column 4 were obtained using the individual experimental values of the differential anomalies. 
In the last two columns, the first uncertainty corresponds to the experiment, and the second corresponds to the theoretical value of the ratio of magnetic anomalies.
}
\centering
\begin{tabular}{lcrr}
\hline
\hline
 Nucleus & Ref.~\cite{barzakh2012hyperfine} & ~~~~~~This work & This work\\
\hline
${}^{187}$Tl$^m$ & 3.707(22) & 3.710(22)(2) & ~~~3.687(38)(2) \\
${}^{189}$Tl$^m$ & 3.756(22) & 3.758(22)(2) & 3.764(42)(2) \\
${}^{191}$Tl$^m$ & 3.781(22) & 3.783(22)(2) & 3.785(24)(2) \\
${}^{193}$Tl$^m$ & 3.824(22) & 3.827(22)(2) & 3.841(25)(2) \\
\hline
\hline
\end{tabular}
\label{table:8}
\end{table}

\section{Conclusion}

In the present paper, we have developed the approach to treat the nuclear magnetization distribution contribution to the hyperfine structure constants in many-electron atoms, which can be used in the calculations with the explicit treatment of the electronic correlation effects. 
The approach can be further generalized to the molecular case.

Using the approach, we have numerically verified that the ratio of the magnetic hyperfine anomalies for a pair of electronic states is rather stable with respect to the choice of the nuclear magnetization distribution model. 
The obtained uncertainty can be taken into account when one uses the ratio for determining the magnetic moments of short-lived nuclei.

It has been demonstrated that the order of magnitude of the differential hyperfine anomaly for Tl isotopes having the $s_{1/2}$ valence nucleon state can be calculated using the model of the uniformly magnetized ball and single-particle models. 
However, the uniformly magnetized ball model cannot be used for isotopes with different nuclear configurations. 
It gives a wrong order of magnitude for the differential hyperfine anomaly. 
At the same time, the single-particle models with a uniform or Woods-Saxon distribution of the valence nucleon give reasonable results.

\section{Acknowledgments}

We are grateful to A. V. Oleinichenko, M. G. Kozlov, A. E. Barzakh, V. M. Shabaev and Yu. A. Demidov for helpful discussions. 
Electronic structure calculations in the paper were carried out using resources of the collective usage center ``Modeling and Predicting Properties of Materials'' at NRC ``Kurchatov Institute'' - PNPI.
The research (except for calculation of the point magnetic dipole HFS constants and Gaunt interaction integrals) has been supported by the Russian Science Foundation Grant No. 19-72-10019. 
Calculations of the point magnetic dipole HFS constants have been supported by the Foundation for the Advancement of Theoretical Physics and Mathematics ``BASIS'' grant according to the Research Project No. 18-1-3-55-1. 
Calculation of the Gaunt contribution has been supported by RFBR Grant No. 20-32-70177.

\end{document}